\documentclass[journal=cmatex,manuscript=article]{achemso}
\usepackage[version=3]{mhchem} 
\usepackage{graphicx}
\usepackage{amsmath}
\usepackage{amssymb}
\usepackage{longtable}

\newcommand*{\citen}[1]{%
  \begingroup%
    \romannumeral-`\x \setcitestyle{numbers}%
    \cite{#1}%
  \endgroup%
}
\newcommand{\beginsupplement}{%
        \setcounter{table}{0}
        \renewcommand{\thetable}{S\arabic{table}}%
        \setcounter{figure}{0}
        \renewcommand{\thefigure}{S\arabic{figure}}%
     }

\title{Predicting the phase stability of multi-component high entropy compounds}

\author{Krishna Chaitanya Pitike}
\email{pitikek@ornl.gov}
\affiliation{%
Materials Science and Technology Division, 
Oak Ridge National Laboratory, Oak Ridge, Tennessee 37831, USA
}%
\author{Santosh KC}
\affiliation{%
Chemical and Materials Engineering,
San Jos\'e State University, San Jos\'e, California 95192, USA
}%
\author{Markus Eisenbach}
\affiliation{%
National Center for Computational Sciences,
Oak Ridge National Laboratory, Oak Ridge, Tennessee 37831, USA
}%
\author{Craig A. Bridges}
\affiliation{%
Chemical Sciences Division, 
Oak Ridge National Laboratory, Oak Ridge, Tennessee 37831, USA
}%
\author{Valentino R. Cooper}
\email{coopervr@ornl.gov}
\affiliation{%
Materials Science and Technology Division, 
Oak Ridge National Laboratory, Oak Ridge, Tennessee 37831, USA
}%

\date{\today}
\begin{document}


\noindent
\textbf{NOTICE:} This manuscript has been authored by UT-Battelle, LLC, under contract DE-AC05-00OR22725 with the U.S. Department of Energy. The United States Government retains and the publisher, by accepting the article for publication, acknowledges that the United States Government retains a non-exclusive, paid-up, irrevocable, world-wide license to publish or reproduce the published form of this manuscript, or allow others to do so, for United States Government purposes. The Department of Energy will provide public access to these results of federally sponsored research in accordance with the DOE Public Access Plan (http://energy.gov/downloads/doe-public-access-plan).
\pagebreak

\begin{abstract}
A generic method to estimate the relative feasibility of formation of high
entropy compounds in a single phase, directly from first principles, is 
developed.
As a first step, the relative formation abilities of 56 multi-component, $A$O, 
oxides were evaluated. These were constructed from 5 cation combinations 
chosen from $A$=\{Ca, Co, Cu, Fe, Mg, Mn, Ni, Zn\}.
Candidates for multi-component oxides are predicted from descriptors related to the enthalpy and 
configurational entropy obtained from the mixing enthalpies of two
component oxides.
The utility of this approach is evaluated by comparing the predicted
combinations with the experimentally realized entropy stabilized oxide, 
(MgCoCuNiZn)O.
In the second step, Monte Carlo simulations are utilized to investigate
the phase composition and local ionic segregation as a function of
temperature.
This approach allows for the evaluation of potential secondary phases, 
thereby making realistic predictions of novel multi-component compounds
that can be synthesized.
\end{abstract}

\maketitle
\section{Introduction
\label{sec:sec1}}

Similar to the explosion in 2D-material systems after the first 
report\cite{Novoselov2004} of graphene preparation by a simple 
micromechanical exfoliation of pyrolytic graphite in 2004, high
entropy material synthesis is emerging as a ``gold rush'' for 
designing multi-component single phase materials.
For example, high entropy (metal) alloys (HEAs) containing 
multi-components (generally $\geq5$) in equal atomic ratios have
been demonstrated to exhibit\cite{Ranganathan2003,Yeh2004,
Cantor2004,Senkov2015,Widom2018} a wide range of remarkable  
mechanical,\cite{Gao2016,Lim2016,Gludovatz2014,Gali2013,Senkov2011,
Li2016,Tsao2017,Li2017,Senkov2010,Senkov2013} 
dielectric,\cite{Berardan2016} and 
superconducting\cite{Rohr2016,Stolze2018} properties. 
In some instances, the single phase multi-component materials inherit
the functional traits of the parent compounds, while in other cases,
they possess emergent properties due to the random/homogeneous 
mixing of five or more components, thus opening avenues for novel 
applications.
High entropy in this context may refer to the thermodynamic
stabilization of a material at finite temperatures, through maximizing 
the entropic contributions to the free energy, $-T{\Delta}S$, rather 
than (or potentially in cooperation with) the mixing enthalpy, 
$\Delta H_\mathrm{mix}$. 
In some cases, high entropy atomic substitution on a crystallographic 
site may not be required to stabilize a material, but may potentially
impart beneficial properties beyond typically observed for the given 
structure type.\cite{Oses2020}
Building upon the success of the HEAs, researchers have attempted to 
synthesize ceramic analogs of the HEAs, such as
high entropy oxides (HEOs). 
(CoCuMgNiZn)O, was the first successful multi-component 
oxide, and was synthesized in a pure rock salt phase.\cite{Rost2015a,Rak2016}
In their seminal work, Rost. et al. demonstrated that an equimolar 
mixture of the parent oxides, CoO, CuO, MgO, NiO and ZnO,
stabilizes in a pure rock salt phase at lower temperatures than any 
closely related off-stoichiometry composition.\cite{Rost2015a}
Since the configurational entropy of the equimolar mixture is
maximum within the ideal mixing model --- correlating with the 
equimolar mixture exhibiting the lowest temperature 
at which a pure rock salt phase is stabilized --- the (CoCuMgNiZn)O 
composition was identified as the first entropy stabilized oxide (ESO).
ESO and related materials with charge compensating cationic 
substitutions, have demonstrated colossal dielectric constant 
which could be exploited for large-k dielectric materials for 
energy storage.\cite{Berardan2016}
These materials are also reported to exhibit large Li‐-ion 
conductivities\cite{Berardan2016a} and potentially large Li‐-storage
capabilities --- both of these properties are highly desirable 
for electrochemical energy conversion in battery 
applications.\cite{Sarkar2018a,Sarkar2019,Wang2019,Loekcue2020}
Entropy stabilization could further benefit storage capacity 
retention and cycling stability in batteries.\cite{Sarkar2018a,Loekcue2020}
Besides energy storage and electrochemical energy conversion, ESO 
have also been reported to exhibit long range antiferromagnetic 
ordering,\cite{JZhang2019,Frandsen2020} and elastic 
anisotropy\cite{Pitike2020}.
Other high entropy perovskite oxides and chalcogenides show potential
photocaltytic hydrogen evolution activity\cite{Edalati2020} and 
semiconducting functionalities,\cite{Deng2020} respectively.
Several other multi-component compounds have since been  
synthesized, such as, high entropy-oxides,\cite{Sharma2018,
Jiang2018,Sarkar2018,Dabrowa2018,Gild2018,Chen2018} 
borides,\cite{Gild2016} and carbides\cite{Castle2018,Dusza2018,
Yan2018,Zhou2018,Sarker2018,Zhang2019a}.
The high entropy oxides were synthesized in several phases with,
one cation sublattice structures such as, rock salt,\cite{Rost2015a} 
and fluorite,\cite{Gild2018,Chen2018} as well as multi-component 
systems with two cation sublattices, e.g. 
perovskites\cite{Sharma2018,Jiang2018,Sarkar2018,Brahlek2020} and 
spinel\cite{Dabrowa2018}.
Despite the fact that several high entropy compounds have been
experimentally realized, there are $>9000$ structure types 
(see the Inorganic Crystal Structure Database (ICSD) Ref. 
[\citen{Bergerhoff1987}]), that could be potentially synthesized 
with five or more cations are present on a single crystallographic
site in equal proportions.
Supplementing the empirical effort, high throughput computations 
could accelerate the exploration of high entropy alloys and compounds.
The significance of high throughput computations in materials
discovery can be emphasized by recent efforts, for example, to 
explore ternary oxides,\cite{Hautier2010} water splitting catalytic 
activity,\cite{Emery2016} battery cathode materials\cite{Hautier2011}, 
predicting phase space,\cite{Ong2008} etc.
Early synthesis attempts of high entropy and entropy stabilized 
materials have predominantly relied on chemical intuition, based on 
relative atomic or ionic radii.
A few previous attempts have utilized theoretical and computational 
methods to predict random and homogeneous mixing of multi-component 
high entropy alloys.\cite{Huang2019,Yang2012,Chattopadhyay2018}
Some theoretical predictions\cite{Troparevsky2015} were limited to cases
where a robust database, was already available, such as for metal
alloys.\cite{Curtarolo2003,Curtarolo2012,Curtarolo2013,Hart2013}
For the cases in which data were not available, 
computational predictions have relied on first principles based 
density functional theory (DFT) computations. 
While such computational efforts have been useful in predicting the
entropy forming ability --- ability/likeliness of a 
multi-component compound to form a single homogeneous phase ---
direct DFT computations suffer from not 
only large computational cost, but also undersampling of the 
configurational landscape prohibited by the size of the supercell.\cite{Sarkar2018a}
In this work, we discuss a generic method, starting from 
first principles, to evaluate the relative feasibility of formation 
of single phase multi-component compounds.
Specifically, we propose enthalpy and entropy descriptors --- used in
quantifying the relative feasibility of formation of single phase 
multi-component compounds --- which can be directly estimated from 
the mixing enthalpies of the two component compounds.
A similar approach was utilized to predict the likeliness of 
formation of HEAs utilizing formation enthalpies of binary 
alloys.\cite{Troparevsky2015}
Here we extend this approach to allow for the prediction of secondary
phases and/or phase segregation of one or more components.
This ability of this framework to discover potential secondary phases,
as well as their tendency to form disordered homogeneous solid solutions 
allows us to make realistic predictions of novel multi-component compounds.
While this work uses the example of divalent multi-component oxides, 
the method can be applied to other multi-component 
materials classes and structure types, such as: alloys, oxides
(perovskites, fluorites, pyrochlores), borides, nitrides, carbides etc.
Development and improvement of these methods is critical for
the computational prediction and targeted experimental exploration of 
novel entropy stabilized materials in generic crystal structure types
that could potentially be exploited in battery applications and other
functionalities such as polar properties in high entropy perovskites.

\section{Computational Methods
\label{sec:sec2}}

\subsection{Details of first-principles calculations
\label{sec:DFT}}
All DFT calculations were carried out using the plane-wave-based Vienna
\textit{Ab-initio} Simulation Package 
\textsc{VASP}\cite{Kresse1996,Kresse1996a}
version 5.4.4, within the Generalized Gradient Approximation
(GGA) using the Perdew-Burke-Ernzerhof for solids (PBEsol) 
exchange-correlation functional.\cite{Perdew2008}
The energy cutoff for the plane-wave basis set was 800 eV, employing
Projected Augmented Wave (PAW) potentials.\cite{Bloechl1994,Kresse1999}
An $8\times8\times8$ $k-$point mesh was utilized for sampling the 
Brillouin zone for a two atom unit cell and scaled linearly with the 
number of atoms present in the unit cell.
The bulk geometry was optimized with a force convergence criterion 
of 1 meV/\AA~ and the individual components of the stress tensor were 
converged to $\leq 0.1$ kB.
Magnetism of Co, Cu, Fe, Mn and Ni oxides were treated with the PBEsol
collinear spin density approximation in the generalized gradient 
approximation, with an onsite Hubbard $U$, (GGA$+U$)
scheme.\cite{Dudarev1998} 
An on‐site Coulomb parameter $U=6$ eV was applied for all cations to 
account for the increased Coulomb repulsion between the semi-filled 
3$d$ states. 
At $U=6$ eV, we obtain a two fold advantage: first, we get the correct
ground state phases for all TMOs considered, and secondly, we get 
insulating states for all pure and two component oxides (TCOs).
We note that, different values between $U=0$ and 8 eV have been used in
the literature to investigate various properties of the TMOs within DFT.\cite{Franchini2005,Fang1999,Cococcioni2005,Boussendel2010,
Schroen2012,Dudarev1998,Deng2010}
The magnetic moments of Co, Fe, Mn and Ni oxides in the rock salt phase 
were initiated in the AFM-II type antiferromagnetic state (shown in 
Figure S1) --- with spins ferromagnetically aligned within
 (111) planes and antiferromagnetically ordered between adjacent (111) planes.\cite{Shull1951,Roth1958}
These structures were also identified to have the AFM-II ground state phase
within DFT and dynamical mean field  theory (DMFT).%
\cite{Franchini2005,Fang1999,Cococcioni2005,Boussendel2010,Schroen2012,Dudarev1998,Deng2010,Zhang2019}

\subsection{Nearest neighbor model
\label{sec:nnModel}}
To computationally explore the configurational energy landscapes, the 
energies of a large number of configurations need to be studied.
Sampling the configurational space directly from DFT calculations is 
computationally expensive and limited by the accessible supercell sizes.
For example, the smallest supercell required to study (CoCuMgNiZn)O in the
rock salt phase --- with highly periodic cation mixing in 
the [1,1,1] crystallographic direction --- contains 10 atoms.
Whereas to obtain reasonable statistics of configurations with random 
cation mixing in three crystallographic directions, 
requires a minimum of 200 atom supercell calculations 
for the rock salt structure.
To reduce the computational costs associated with directly 
studying a five component oxide (FCO), we adopt a nearest neighbor 
model, (NNM) whose parameters can be relatively easily obtained from
two component oxides.
Essentially, the NNM defines the local energy of a cation by regarding 
all first nearest neighbor cations as independently interacting ions.
In this picture, the interactions are limited to the first nearest 
neighbor cations, and solely dependent upon the composition
of the first nearest neighbor cation coordination shell.
It is worth noting that the interactions between the first nearest 
neighbor cations are mediated through the interpenetrating oxygen 
sublattice. 
The mixing enthalpies between two cations are obtained by relaxing the
atomic models within DFT.
Hence the mixing enthalpies of two cations implicitly include the ionic
relaxation of both cation and oxygen sublattices, arising due to the 
differences in the ionic radii of the cations.
The enthalpies of mixing, $\Delta H_\mathrm{mix}$ between two cations 
in a generic phase are obtained from the DFT optimized total energies of
the rock salt ordered two component oxide structures.
Further information on these energetics is found below in Sec III.A.
The NNM is a simple yet powerful tool to explore the potential of a 
composition to form ordered versus disordered structures and has been
successfully utilized to study
HEAs.\cite{Hillert1962,Troparevsky2015,Santodonato2018}
Here we employ the NNM as a basis for studying formation ability  
and local segregation in multi-component oxides.
\section{Results and Discussion
\label{sec:sec3}}
\subsection{Structure and enthalpies of mixing
\label{sec:structure}}

\begin{figure}[!htbp]
\includegraphics[width=0.5\columnwidth]{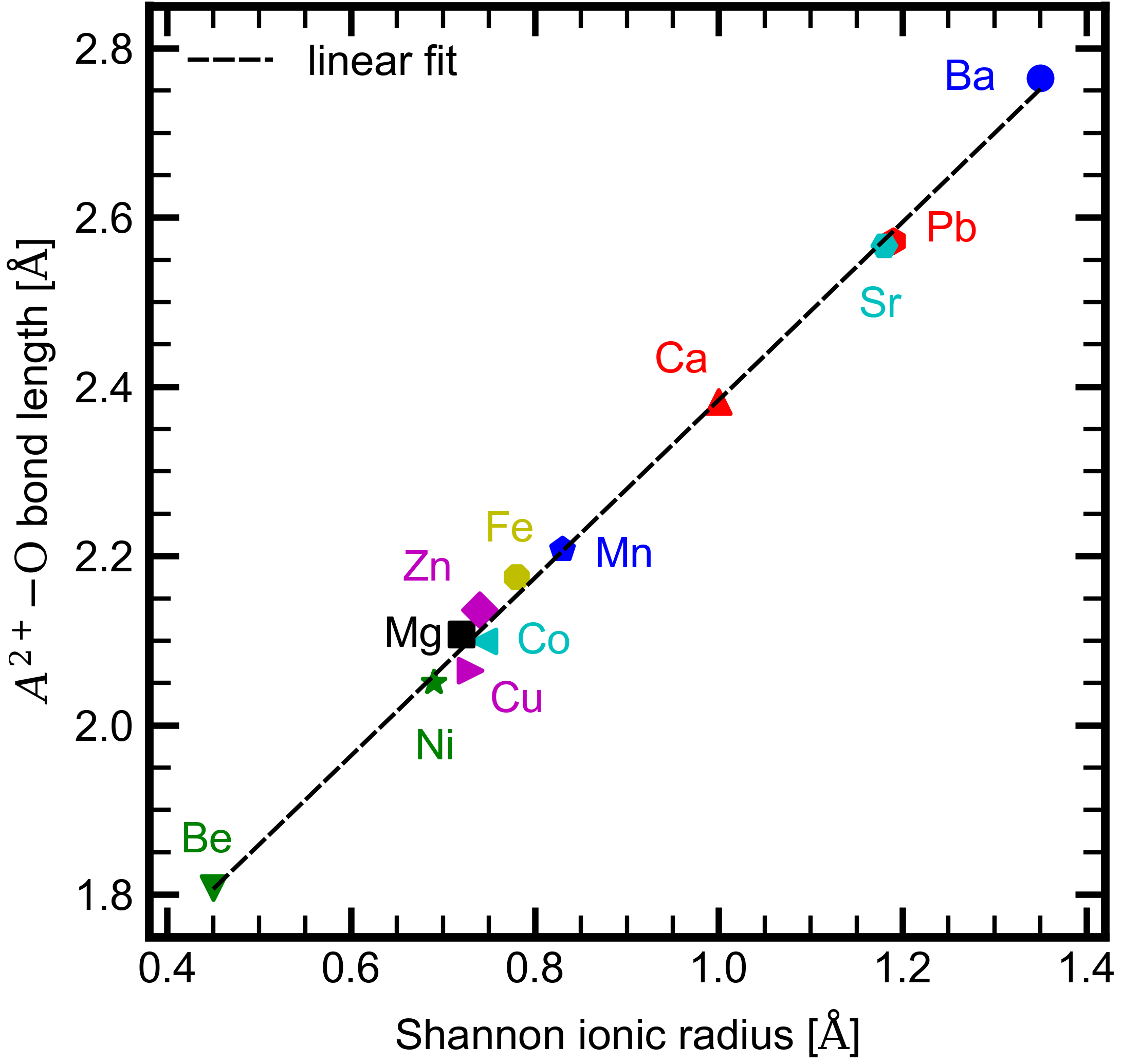}
\caption{\label{fig:radii}
The Shannon ionic radii is plotted against the respective DFT computed
$A^{2+}-\mathrm{O}$ bond length for several divalent cations in the rock salt
phase.
The cations constituting the original entropy stabilized oxide 
--- Co, Cu, Mg, Ni and Zn --- are clustered at the bottom of the plot, 
along with Fe and Mn.
Where as other cations --- Ca, Sr, Ba and Pb --- have large ionic radii.
}
\end{figure}
Figure \ref{fig:radii} presents the $A-$O bond length in the rock salt 
phase as a function of $A$ cation Shannon ionic radii \cite{Shannon1976}
for most common divalent cations --- Be, Ba, Ca, Co, Cu, Fe, Mg, Mn, Ni,
Pb, Sr and Zn.
As expected, the $A-$O bond lengths of divalent cations have strong
linear correlation with their Shannon ionic radii.
Interestingly we observe that the Shannon ionic radii of the cations 
comprising the original ESO --- Co, Cu, Mg, Ni and Zn --- are within
0.055 \AA\ of each other.
Chemical intuition would suggest that mixing cations with very 
different Shannon radii would be more likely to result 
in either phase segregation, or a different structure type, or the 
smaller cation occupying interstitial sites.
Based on the $A-$O bond lengths (or Shannon ionic radii), in the 
remaining analysis, we only consider the following eight cations 
--- Ca, Co, Cu, Fe, Mg, Mn, Ni, Zn.
From these 8 cations, a total of $8!/5!3!=56$ five cation compositions
can be generated.
These 56 cations combinations along with their index are given in the 
supplemental information.
While CaO, CoO, MgO, and NiO readily crystallize in the rock salt
structure, their ground state, CuO and ZnO are stable in the tenorite
and wurtzite structures, respectively.
While, Fe and Mn form stable Fe$_2$O$_3$ and Mn$_2$O$_3$ their 
divalent oxides, FeO and MnO, can be experimentally realized at  
room temperature and ambient pressure in the rock salt 
structure.\cite{Bergerhoff1987,Katsura1967,SASAKI1979}
As an initial step to understanding the tendency to form a single 
phase solid solution we first evaluate the structure and mixing 
enthalpies of two component oxides, $(AA^{'})O$.
A total of 28 TCO compositions can be generated using the 8 divalent 
cations of interest ($8!/6!2! = 28$).
From the experimental literature, we find that the TCOs may have ground state
phases in one of four structures --- rock salt, tenorite, wurtzite
and zinc blende.
Hence, in addition to the three structure found in the parent oxides, 
i.e., rock salt, tenorite and wurtzite, we also consider the zinc 
blende structure for further analysis.
Structural models of the four structures are included in the 
supplemental material.
The mixing enthalpy, $\Delta H_\mathrm{mix}\big[(AA^{'})\mathrm{O},
P\big]$, of $(AA^{'})$O TCO in a generic phase, $P$, is estimated 
from DFT calculations as shown in Eq. (\ref{eq:ME}).
\begin{align}
\begin{split}
\label{eq:ME}
\Delta H_\mathrm{mix}\big[(AA^{'})\mathrm{O},P\big]{}&=
E_\mathrm{DFT}\big[(AA^{'})\mathrm{O},P\big]
-\frac{1}{2m}E_\mathrm{DFT}\big[A_m\mathrm{O}_x,\mathrm{G}\big]\\
-&\frac{1}{2n}E_\mathrm{DFT}\big[A^{'}_n\mathrm{O}_y,\mathrm{G}\big]
-\frac{1}{2}(1-\frac{x}{2m}-\frac{y}{2n}) 
E_\mathrm{DFT}\big[\mathrm{O_2},\mathrm{G}\big]
\end{split}
\end{align}
where, $E_\mathrm{DFT}\big[A_m\mathrm{O}_x,\mathrm{G}\big]$, 
$E_\mathrm{DFT}\big[A_n^{'}\mathrm{O}_y,\mathrm{G}\big]$ 
are the DFT total energies of the $A_m$O$_x$ and $A_n^{'}$O$_y$ 
oxides in their respective ground state phases.
$E_\mathrm{DFT}\big[(AA^{'})\mathrm{O},P\big]$ is the DFT total 
energy of the $(AA^{'})$O TCO in phase $P$.
In the spirit of the NNM, the value of 
$E_\mathrm{DFT}\big[(AA^{'})\mathrm{O},P\big]$,
in its ground state phase, is used to describe the interaction energy 
between $A-A^{'}$ cations, $\Delta H_\mathrm{mix}^{A-A^{'}}=
\Delta H_\mathrm{mix}\big[ (AA^{'})\mathrm{O},G \big]$ in the FCO.
\begin{figure*}[!htbp]
\includegraphics[width=1\columnwidth]{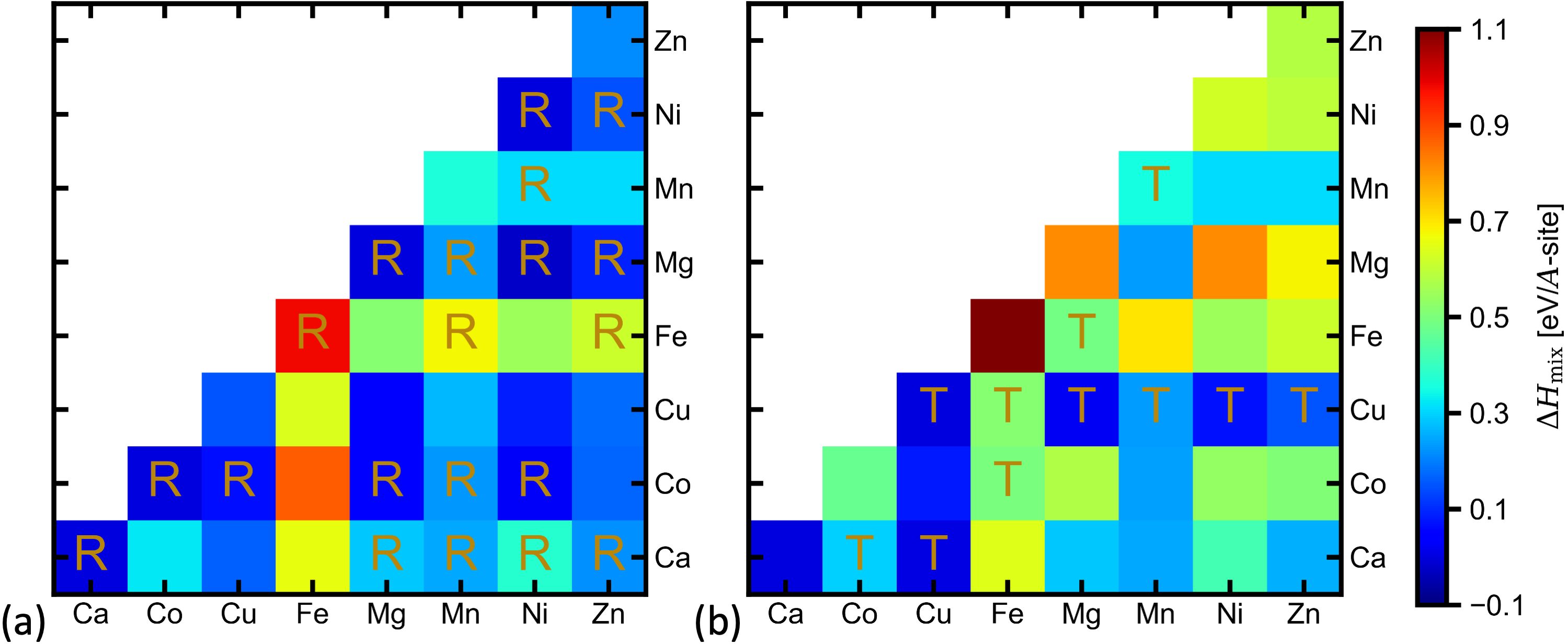}
\caption{\label{fig:ME}
Heat maps for mixing enthalpy in eV/$A$-site of $(AA^{'})$O two component 
oxides in phase $P$, $\Delta H_\mathrm{mix}((AA^{'})\mathrm{O},P)$,
calculated from first principles using Eq.~(\ref{eq:ME}).
\textbf{(a)} and \textbf{(b)} shows the mixing enthalpy 
for rock salt and tenorite phases, respectively.
The annotations for the rock salt and tenorite ground phases for each
combination are represented with R and T, respectively.
Only the lower triangular portion of the symmetric data is shown in all 
panels to avoid confusion.
The color legend represents the mixing enthalpy in [eV$/A$-site] --- 
blue and red represents low ($\sim$ -0.1) and high ($\sim$ 1.1) bond 
energies, respectively.
}
\end{figure*}

Figure \ref{fig:ME} presents the mixing enthalpy heat maps calculated
using Eq.~(\ref{eq:ME}).
The mixing enthalpies for the rock salt, tenorite and ground phases are 
shown in Figure \ref{fig:ME}(a) and (b), respectively.
Similar plots for the minimum energy phase, wurtzite and zinc blende 
phases are shown in the supplemental material.
Out of these 28 unique TCO combinations and 8 parent oxides (represented
by the diagonal elements), we find that there are 20, 11, 3 and 2 
compounds with rock salt, tenorite, wurtzite and zinc blende minimum 
energy structures, respectively.
Here we would like to point out that we have only considered two 
component oxides with equiatomic mixing in four different structures.
Furthermore, a two component oxide with positive minimum energy is 
thermodynamically unstable and in such cases the two component combination 
is more likely to form a two-phase system, either due to a lack of reaction 
or only a partial reaction between the components.
\subsection{Enthalpy and entropy descriptors from mixing enthalpies
\label{sec:descriptors}}
\begin{figure}[!htbp]
\includegraphics[width=0.5\columnwidth]{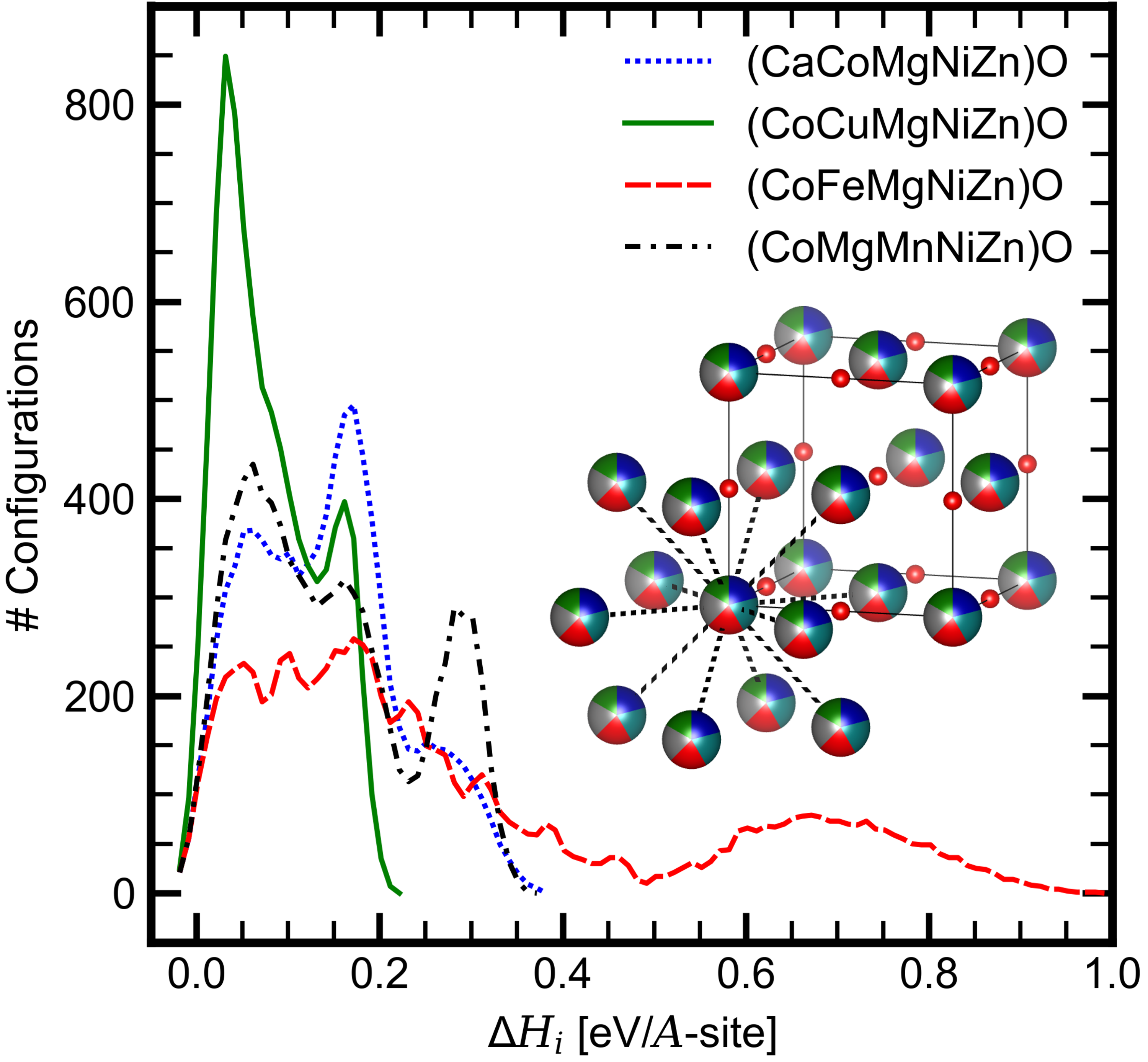}
\caption{\label{fig:local}
The energy distribution of the local mixing enthalpies of all possible 
configurations are plotted for four FCO combinations in rock salt phase.
The solid green line shows the distribution for (CoCuMgNiZn)O. 
Three additional representative combinations --- obtained by replacing 
Cu with Ca, Fe and Mg --- are plotted in blue dotted, red dashed and 
black dot-dashed lines, respectively.
}
\end{figure}

The descriptors for the enthalpy and entropy, contributing to the free 
energy, are obtained from exploring the local mixing enthalpies 
within the five component oxide.
In the NNM framework, the expectation value of the local mixing
enthalpy of a five component oxide, $\langle \Delta H_\mathrm{local} 
\rangle$, is obtained from exploring all local configurations.
There are twelve first nearest neighbor cations surrounding each cation
in the rock salt structure.
A local atomic configuration at a lattice site, $i$, constitutes 
chemical species occupied by the lattice site, $A^i$, along with the 
twelve neighboring species, $A^{j}$, $j=\{1,2,...12\}$.
This local atomic configuration is presented in the inset of Figure 
\ref{fig:local}.
The total number of unique local configurations are obtained by 
iterating through the combinations of ($r=5$) five chemically 
distinct cations with replacement in ($n=12$) twelve surrounding 
$A$-sites, $N_\mathrm{local}=n \times 
^{(n+r-1)}\mathrm{C}_{(r-1)}=\frac{(n+r-1)!}{(n-1)!(r-1)!}=9100$.
The local mixing enthalpy in the local phase $P_i$ is estimated 
through,
\begin{align}
\begin{split}
\label{eq:LME1}
\Delta H_i\big[\mathrm{FCO},P_i\big]=\frac{1}{n}\sum\limits_{j=1}^n 
\Delta H_\mathrm{mix}\big[(A^iA^j)\mathrm{O},P_i\big]
\end{split}
\end{align}
where, $\Delta H_\mathrm{mix}\big[(A^iA^j)\mathrm{O},P_i\big]$ is 
the average mixing enthalpy of $A^i$ with an $A^j$  
cation species, computed from Eq.~(\ref{eq:ME}).

Figure \ref{fig:local} plots the distribution of local mixing 
enthalpies, $\Delta H_i$, in the rock salt phase for the original ESO 
and three representative FCOs obtained by replacing Cu by Ca, Fe and Mn.
The $\Delta H_i$ distribution for all 56 FCOs are shown in the 
supplemental material.
The distribution of the local mixing enthalpies suggests that the 
number of local configurations accessible through 
thermal excitations are highest for the (CoCuMgNiZn)O, represented 
by the narrow green peak between $E_\mathrm{local}=$ 0.0 and 0.2 eV.
We note that the configurational entropy contribution to the Gibbs 
free energy should be higher for a FCO combination with a narrow
distribution of the local mixing enthalpies.
The configurational entropy is inversely related to the 
number of configurations accessible through the thermal excitations.
Further on this will be elaborated later in this section.
Here, we introduce a set of descriptors that quantify the enthalpy and 
configurational entropy for a FCO.
These are obtained from the mean, $\mu_\mathrm{local}$, and 
standard deviation, $\sigma_\mathrm{local}$, of the local mixing 
enthalpy distributions defined as:
\begin{subequations}
\label{eq:localstats}
\begin{align}
\mu_\mathrm{local}                 = {}& 
\langle \Delta H_\mathrm{local} \rangle ~~\mathrm{and} \\
\sigma_\mathrm{local}              = {}&
\sqrt{\langle \Delta H_\mathrm{local}^2 \rangle - \langle \Delta H_\mathrm{local} \rangle^2} ~~\mathrm{;}\\
\langle \Delta H_\mathrm{local} \rangle   = {}& 
\frac{1}{N_\mathrm{local}} \sum_{i=1}^{N_\mathrm{local}} \Big(\Delta H_i\big[\mathrm{FCO},P_i\big]\Big) ~~\mathrm{and}\\
\langle \Delta H_\mathrm{local}^2 \rangle = {}&
\frac{1}{N_\mathrm{local}} \sum_{i=1}^{N_\mathrm{local}} \Big(\Delta H_i\big[\mathrm{FCO},P_i\big]\Big)^2 ~~\mathrm{;}
\end{align}
\end{subequations}
where, $\langle \Delta H_\mathrm{local} \rangle$ and 
$\langle \Delta H_\mathrm{local}^2 \rangle$ are the mean values of 
the local mixing enthalpies and squared local mixing enthalpies,
respectively.
The significance of $\mu_\mathrm{local}$ and $\sigma_\mathrm{local}$
can be explained by considering the following two FCOs, with, 
(1) low-$\mu_\mathrm{local}$ and low-$\sigma_\mathrm{local}$,
(2) high-$\mu_\mathrm{local}$ and high-$\sigma_\mathrm{local}$.
For the purpose of this discussion we restrict ourselves to the case 
where $\mu_\mathrm{local}$ is positive, therefore indicating a 
preference for phase segregation, the case of $\mu_\mathrm{local}$ 
will be discussed later. 
In the first case, the small but positive mixing enthalpy 
contribution to the Gibbs free energy indicates that although the
compound would prefer not to form a solid solution it may be possibly
overcome by entropy. 
Commensurately, if the distribution of configurational states is 
relatively narrow, i.e. $\sigma_\mathrm{local}$ is small, then the 
entropy contribution (which is inversely proportional to 
$\sigma_\mathrm{local}$) would be large. 
Here, the contributions from configurational entropy are largely due
to the large number of degenerate local configurations, accessible 
through thermal excitations. As such, a pure single phase structure
will be formed at $T=T_t$ where the entropy contribution overcomes 
the small and positive enthalpy contribution. 
Such a case is exemplified in the the distribution of configurations 
for (CoCuMgNiZn)O as depicted in Fig.~\ref{fig:local} (red dashed line).
Conversely, large $\mu_\mathrm{local}$ is indicative of a strong 
preference for phase segregation. 
However, it can be shown that in all possible FCOs from obtained by 
mixing 8 parent oxides, there will be at least two cations with 
favorable mixing enthalpies. 
This is suggestive of a disperse distribution of mixing enthalpies. 
Such a diffuse/distributed configurational landscape, would result
in a very large $\sigma_\mathrm{local}$ contributing to a small 
configurational entropy, which will be insufficient to compensate 
for the already large $\mu_\mathrm{local}$. 
This behavior is demonstrated in the energy distribution of 
(CoFeMgNiZn)O depicted in Fig.~\ref{fig:local}.
This interpretation indicates that the values of both the enthalpy 
and entropy descriptors, $\mu_\mathrm{local}$ and $\sigma_\mathrm{local}$, 
respectively, need to be considered for predictions of formation ability
of a multi-component, single phase structure.
We propose that an FCO with low-$\mu_\mathrm{local}$ and 
low-$\sigma_\mathrm{local}$, would be a prime candidate for forming a 
multi-component, entropy-stabilized single phase material.
\begin{figure}[!htbp]
\includegraphics[width=0.5\columnwidth]{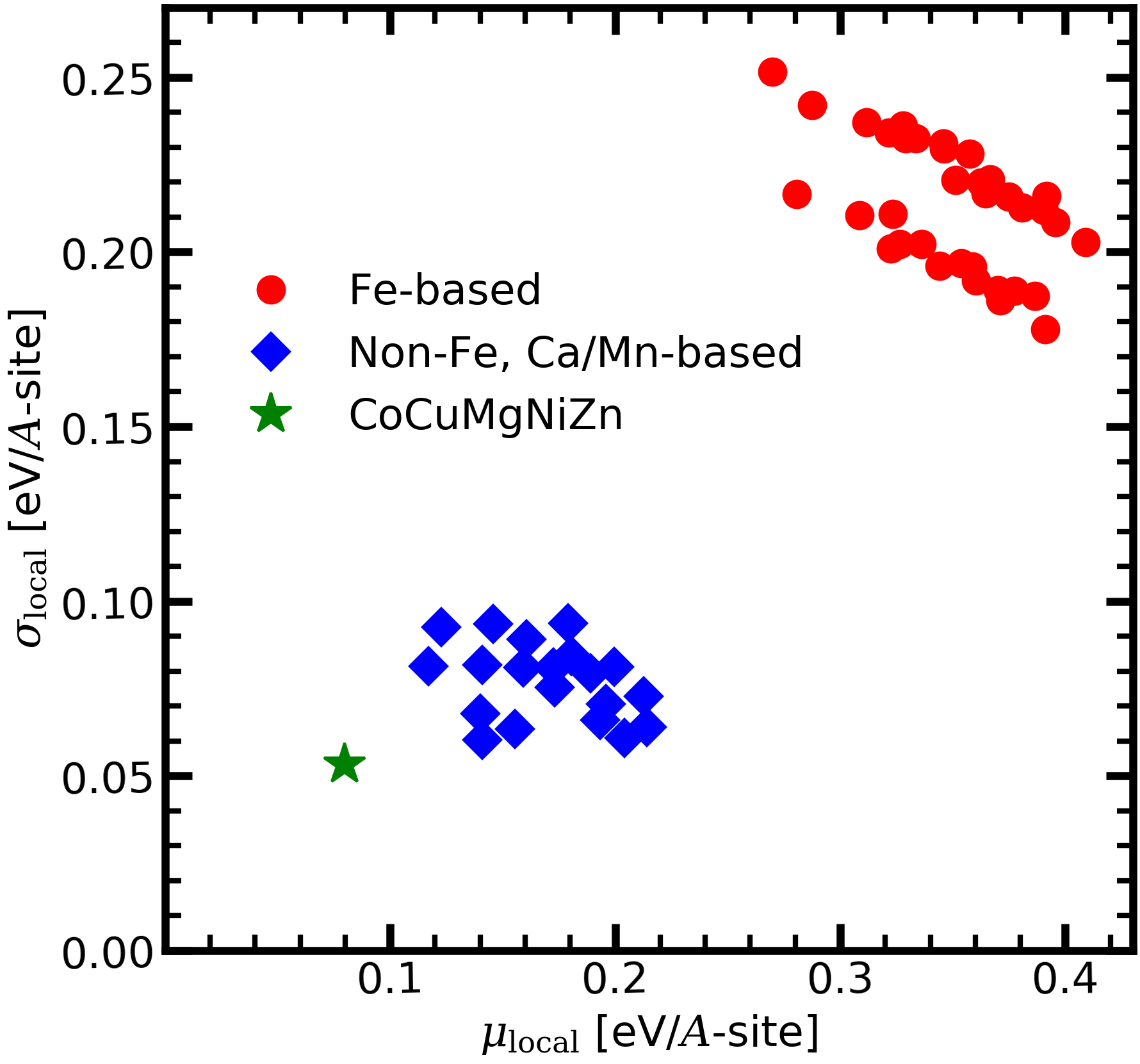}
\caption{\label{fig:mul-sigmal}
A comparison of the enthalpy and entropy descriptors, $\mu_\mathrm{local}$ and 
$\sigma_\mathrm{local}$, respectively, for all 56 FCOs.
(1) Fe-based FCO cluster is shown in red, 
(2) Ca- or Mn-based (without Fe) FCO cluster is shown in blue, and 
(3) The entropy stabilized oxide --- (CoCuMgNiZn)O --- shown in green, is unique FCO
with least-$\mu_\mathrm{local}$ and least-$\sigma_\mathrm{local}$. 
}
\end{figure}
Figure \ref{fig:mul-sigmal} presents a comparison of the enthalpy and 
entropy descriptors, $\mu_\mathrm{local}$ and $\sigma_\mathrm{local}$,
respectively, for all 56 FCOs.
Surprisingly, the FCOs fall into three clusters identified by red
circles, blue diamonds and a green star.
First, remarkably in agreement with the experimental realization, 
our enthalpy and entropy descriptors, predict that the FCO 
represented by the green star, is distinct with  $\mu_\mathrm{local}$
and $\sigma_\mathrm{local}$ close to 0; thereby having the highest 
potential for forming a uniform rock salt phase.
Second, the 20 FCOs in the blue cluster, obtained by replacing one
or two cations in the ESO with either Ca or Mn, have  
$\mu_\mathrm{local}$ and $\sigma_\mathrm{local}$ that range between
$(0.12, 0.22)$ eV and $(0.06, 0.10)$ eV, respectively.
On average the enthalpy contribution is higher than the ESO with a 
lower number of degenerate low-enthalpy configurations.
The above two conditions suggests that, even large temperatures might
not be sufficient for entropy to overcome the large 
unfavourable enthalpy.
For example, the (CaCoCuMgZn)O FCO, located at the bottom of the blue 
cluster, has the second lowest standard deviation, only 
$\Delta \sigma_\mathrm{local} \sim 7$ meV/\AA~ greater than the ESO.
Conversely, the mixing enthalpy of this composition is significantly
$\Delta \mu_\mathrm{local} \sim 60$ meV/\AA~ greater than the ESO.
Hence, by comparing both entropy and enthalpy descriptors, we get a 
comprehensive picture of the formation ability and an indication of
the temperature required to stabilize a single phase multi-component
compound.
We find a third cluster, shown in red circles, formed from the 
remaining 35 Fe-based FCOs, whose values of $\mu_\mathrm{local}$ and 
$\sigma_\mathrm{local}$ are on the order of 0.35 eV and 0.25 eV, 
respectively.
This indicates an extremely unfavourable mixing of Fe within the rock 
salt structure.
The large unfavourable mixing enthalpies exhibited by Fe based FCOs
--- due to the strong preference of Fe to form Fe$_2$O$_3$ rather than 
FeO --- shifts the enthalpies of the local configurations up 
to 1 eV/$A-$site (shown in representative red curve in Figure
\ref{fig:local}).
As discussed above, FCOs containing such configurations with large 
mixing enthalpies --- would not be accessible to thermal excitations 
under the physical temperatures used for solid state reactions.
The large variation in the enthalpies of Fe based FCOs results in lower
configurational entropy.
Similar to Fe, Mn based compounds exhibit unfavorable mixing in the 
rock slat phase due to its preference to form trivalent oxide, albeit 
the tendency is not strong as Fe.
From this enthalpy and entropy descriptors, we find that the 
experimentally realized (CoCuMgNiZn)O is perhaps the only candidate
that can form a homogeneous rock salt phase.
Since the enthalpy of mixing is positive for (CoCuMgNiZn)O, the 
structure must be stabilized through the configurational 
entropy contributions, i.e. it is an entropy stabilized oxide.
Following the above analysis, we possibility of forming alternate 
structures, i.e. tenorite, wurtzite and zinc blende.
Comparisons of the enthalpy and entropy descriptors, for all 56 FCOs
in the tenorite, wurtzite and zinc blende structures are shown in 
the supplemental material.
In all cases, we find  the enthalpy and entropy descriptors for the 
56 FCOs in these phases to be large and positive. 
Thus it is highly unlikely that any binary structures considered other 
than rock salt, will be stabilized through entropy at experimentally 
accessible temperatures.
As an aside, although negative $\mu_\mathrm{local}$ was not observed
for the considered FCOs, we can expect that  multi-component compounds
with negative mixing enthalpies exist. 
One scenario is that a multi-component compound may readily form a pure
phase due to a large preference of all the ions to form that common phase.
Such is the case for many multi-component 
perovskites.\cite{Muromachi1988,Brahlek2020,Sharma2018}
Hence in this case, the configurational entropy is not critical for the
stability of the single phase.
These multi-component compounds with negative mixing enthalpies cannot
be considered entropy stabilized compounds.
Here, we discuss the advantages of utilizing the mixing enthalpies
of TCOs to obtain entropy and enthalpy descriptors compared with
alternate frameworks that were previously proposed.
Sarker et. al. [\citen{Sarkar2018}], proposed a similar
approach based only on an entropy descriptor, derived from the standard 
deviation of the energy distribution
estimated through DFT calculations on a small supercell. 
In their approach, the significance of the mixing enthalpy contribution
--- indicating the energetic favorability of mixing, which correlates 
with the temperature required for obtaining a pure phase --- 
was not considered.
For example, an FCO with a small $\sigma_\mathrm{local}$ could still 
have a large mixing enthalpy whose configurations are inaccessible 
through thermal excitations at physically realizable temperatures; 
as is the case for the (CaCoCuMgZn)O FCO discussed above. 
Using this approach you would have an entropy forming ability
(defined as $1/\sigma_\mathrm{local}$) ranging from 8 to 38
(eV/atom)$^{-1}$; with 38 and 33 (eV/atom)$^{-1}$ for the ESO 
and (CaCoCuMgZn)O, respectively. This would suggest formability
of (CaCoCuMgZn)O without consideration of the large 
enthalpy of formation that would need to be overcome.
In alternate work by Troparevsky et. al. [\citen{Troparevsky2015}], 
a similar approach to Ref. [\citen{Sarkar2018}], was exercised through
a qualitative comparison of the ``closeness'' of the mixing enthalpies 
of binary metal alloys.
Our current approach undertakes a more comprehensive analysis 
by computing the mean and standard deviation of the local
configurational landscapes, related to the enthalpy and entropy 
contributions.
Nevertheless, using their approach  --- obtained 
from the mean, $\mu_\mathrm{mix}$, and standard deviation, 
$\sigma_\mathrm{mix}$, of the mixing enthalpies of 15 two
component combination, as shown below:
\begin{align}
\begin{split}
\label{eq:simplestats}
\mu_\mathrm{mix} = &{} \big\langle \Delta H_\mathrm{mix}[(AA^{'})\mathrm{O},P] \big\rangle \\
\sigma_\mathrm{mix} = &{} \sqrt{ \big\langle \Delta H_\mathrm{mix}[(AA^{'})\mathrm{O},P]^2 \big\rangle - \big\langle \Delta H_\mathrm{mix}[(AA^{'})\mathrm{O},P] \big\rangle^2}
\end{split}
\end{align}
we find qualitative agreement for all 56 FCOs in the rock salt phases, 
with the ordering in Figure \ref{fig:mul-sigmal}.
\subsection{Estimating the phase transition temperature through Metropolis Monte Carlo
\label{sec:mcResults}}
\begin{figure*}[!htbp]
\includegraphics[width=1\columnwidth]{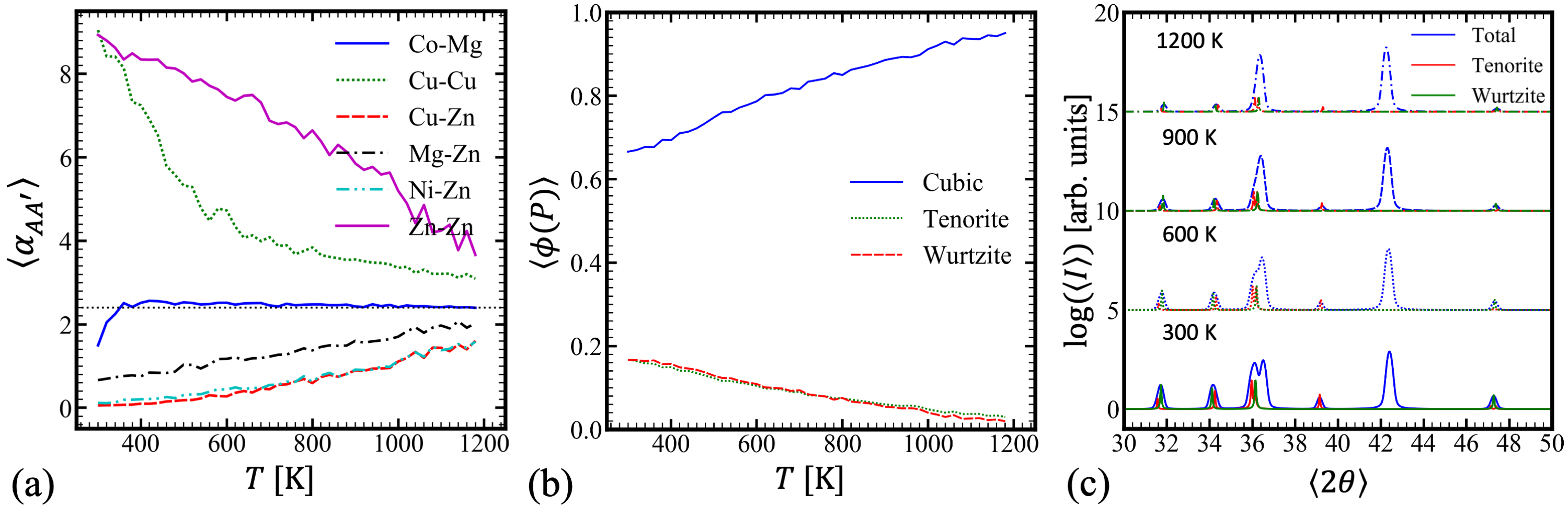}
\caption{\label{fig:montecarlo}
Shows the variation of \textbf{(a)} short range order parameter, $\alpha_{AA^{'}}$,
and \textbf{(b)} phase fraction, $\phi(P)$.
\textbf{(a)} Six of the fifteen unique $\alpha_{AA^{'}}$ parameters
are shown.
All $\alpha_{AA^{'}}$ parameters are shown in supplemental material.
\textbf{(b)} The phase fraction is shown for rock salt, tenorite and 
wurtzite phase.
The phase fraction of rock salt phase, $\phi(P=\mathrm{C})$, increases
with $T$, at the expense of both tenorite and wurtzite phases.
\textbf{(c)} Simulated x-ray diffraction peaks, obtained from the
phase fraction are shown at four differnet temperatures.
}
\end{figure*}
The enthalpy and entropy-related descriptors, discussed above, 
are the first step in predicting the relative likeliness of formation
of an entropy stabilized phase.
In this section, we assess the temperature evolution of the phase 
composition and cation segregation in the FCOs, specifically in,
(CoCuMgNiZn)O utilizing Metropolis Monte Carlo simulations on a 
5$\times$5$\times$20 supercell that contains 2000 cations.
A detailed description of the Monte Carlo method, configurational and
phase composition sampling is given in the supplemental material.
Briefly, the total mixing enthalpy of the FCO lattice model microstate,
in the global phase $P$, $\Delta H_\mathrm{total}[\mathrm{FCO},P]$, is 
estimated by summing over the local mixing enthalpies of all $N=2000$ 
cation lattice sites,
\begin{align}
\begin{split}
\label{eq:TE}
\Delta H_\mathrm{total} \big[ \mathrm{FCO},P \big] = 
\sum_{i=1}^N \Delta H_i \big[ \mathrm{FCO},P_i \big]
\end{split}
\end{align}
$\Delta H_i[\mathrm{FCO},P_i]$ is the local mixing enthalpy at lattice
site $i$, in the local phase $P_i$ as described in Eq. (\ref{eq:LME1}). 
The phase composition of each microstate is achieved by assigning a 
local ground state phase, $G_i$, to each lattice site $i$ as:
\begin{align}
\begin{split}
\label{eq:ground}
\Delta H_i \big[ \mathrm{FCO},G_i \big] = \mathrm{Min} \Big( \Delta H_i \big[ \mathrm{FCO},P_i \big] \Big)
\end{split}
\end{align}
It is crucial to note that the ground state phase, $G$, of a 
microstate, is a combination of all local phases.
For example, at lattice site $i$, a chemical species $A^i$ is 
assigned a local phase $G_i$ --- which is determined based on the 
minimum of the mixing enthalpies of all considered phases.
Defining the total energy according to Eq. (\ref{eq:ground}), allows 
for the estimation of phase fraction, $\phi(P)$, of a phase $P$, by 
dividing the number of occurrences of local phase $P_i$, by the total
number of lattice sites.
\begin{align}
\begin{split}
\label{eq:PF}
\phi(P) = \frac{1}{N}\sum_{i=1}^N \delta(G_i,P)
\end{split}
\end{align}
where, $\delta(G_i,P)$ is a Dirac delta function, which returns 1 
if the local ground state phase at lattice site $i$, $G_i=P$.
While the phase fraction, $\phi(P)$, represents the phase composition
of the microstate, it does not provide any information about local
clustering of one or more cations.
Any local short range ordering, representing a precipitate/secondary
phase can be  estimated by a short range order parameter, 
$\alpha_{AA^{'}}$.
Within the nearest neighbor model, $\alpha_{AA^{'}}$ can be easily 
estimated from a microstate by averaging the number of atoms of species
$A^{'}$ surrounding all atoms of species $A$.
By the commutative relation, $\alpha_{AA^{'}}=\alpha_{A^{'}A}$,
there are only fifteen unique short range order parameters in a five
component high entropy compound.
Figure \ref{fig:montecarlo}(a) depicts the variation of 
$\langle \alpha_{AA^{'}} \rangle$ as a function of $T$.
Essentially, $\langle \alpha_{AA^{'}} \rangle$ represents the likelihood
of finding A and A' associated with each other (if they prefer to be
together, $\langle \alpha_{AA^{'}} \rangle$ will be larger).
Here we focus on 6 of the 15 possible $A-A^{'}$ combinations, 
as shown in Figure \ref{fig:montecarlo}(a).
A more complete picture is given in Figure S6.
For Mg$-$Co, we see that $\langle \alpha_\mathrm{CoMg} 
\rangle$ quickly converges to $\sim 2.4$ at $T=400$ K.
This is the expected value for a perfectly
random FCO $=\frac{12}{5}=2.4$ atoms.
The reader should note that, due to finite size
effects, we consider ideal random mixing to be achieved when 
$2 \leq \langle \alpha_{AA^{'}}\rangle \leq 3$.
Contrarily, we observe that Zn prefers to strongly segregate with 
itself, evident from $\langle \alpha_{AA^{'}} \rangle$  
decreasing to only $\sim 4.0$ at $T \approx 1140$ K. 
Other crucial observations related to each chemical species relevant
to the high temperature evolution are given below.
\begin{enumerate}
    \item Zn prefers to segregate and avoids mixing with Cu and Ni.
    At $T \approx 1140$ K, Zn can accommodate $\frac{2}{3}$ of its 
    neighbors with other cations, typically, Mg and Co.
    \item  Cu avoids mixing with Zn and has a slight tendency
    to segregate. Cu can accommodate mixing with other cations above 
    $\sim 700$ K.
    \item Co, Mg and Ni readily form disordered solid solutions 
    --- perhaps owing to their shared rock salt structure. 
    First nearest neighbor interactions between the cations
    can incorporate Zn and Cu at higher temperatures.
\end{enumerate}
Figure \ref{fig:montecarlo}(b) illustrates the variation of 
$\langle \phi(P) \rangle$ as a function of $T$.
We find that the phase fraction of the tenorite and wurtzite phases
at low temperatures $T \sim$ 300 K is $ \approx 0.2$.
The tenorite and wurtzite phases, occur in Cu and Zn rich 
regions of the microstate, and are indicative of their atomic 
fraction.
Similarly, the phase fraction of the rock salt phase, at lower 
temperatures $ \approx 0.6$, corresponding to the Mg, Co and Ni 
rich regions of the microstate, also agree with their combined
atomic fraction.
As the rock salt phase fraction increases to $\approx 0.95$ at 
$T \approx 1200$ K, the tenorite and wurtzite phase fractions
simultaneously decrease to $\approx 0.03$ and $0.02$, respectively.
These values of the phase fractions indicate that small amounts of 
tenorite and wurtzite phases persists even at large temperatures.
Figure \ref{fig:montecarlo}(c) shows the simulated X-ray diffraction 
patterns obtained by a simple linear combination of intensities of 
the phases present at $T=$ 300, 600, 900 and 1200 K. 
The coefficients for the linear combination at each temperature are
obtained from the phase fraction, $\phi(P)$.
(The intensities are plotted on a log scale 
similar to Figure 1 in the supplemental information of 
Ref. [\citen{Rost2015a}]).
Since the simulated X-ray intensities are free from noise, the 
tenorite$+$wurtzite oxygen sublattice peaks, appearing at 
$2\theta=31.6^\circ$ persevere even at 1200 K.
Nevertheless, the relative intensities of the peaks are diminished by 
$\sim4$ and $\sim6$ times as the temperature is increased to 900 and 
1200 K, respectively.
Interestingly we find that, with the exception of the small diffraction
peak near $2\theta=38.5^\circ$, the wurtzite phase peaks overlap with 
the tenorite phase peaks.
Due to this overlap, it could be quite difficult to identify the 
wurtzite phase through X-ray diffraction alone.
Nevertheless, our nearest neighbor model predicts that at low  
temperatures, the tenorite and wurtzite secondary phases coexist; 
decreasing with increasing temperature.
This agrees well with experimentally observed phase transition
from a mixed phase to a high temperature single phase.\cite{Rost2015a}

\section{Conclusions}
In conclusion, we have developed a generic framework to predict high 
entropy materials and their potential secondary phases as a function 
of temperature.
We achieve this by identifying, simple descriptors, to quantify  
the relative feasibility of formation of single phase high entropy
materials.
These descriptors are obtained from the statistical mean and 
standard deviations of local mixing enthalpies of the five component
oxides.
The local mixing enthalpies are computed from the mixing enthalpies
of two component oxides, within the first nearest neighbor model 
framework, utilizing DFT calculations.
In this picture, the mean is related to the mixing enthalpy of the 
FCO, and the standard deviation is inversely related to the
configurational entropy.
Hence a multi-component system with low, positive, mean (low mixing 
enthalpy) and low standard deviation (high configurational entropy), 
should readily form a single phase high entropy compound.
The proposed descriptors correctly identify the experimentally
realized ESO, (CoCuMgNiZn)O, as the composition most likely to form 
a high entropy rock salt phase.
Building upon this model, we simulate the tendency of one or more 
components to segregate as a function
of temperature through Metropolis Monte Carlo simulations.
This allows us to examine phase compositions as a function of temperature.
In the original ESO, we find that while Mg, Co and Ni readily form 
disordered solid solutions in the rock salt phase, Cu an Zn prefer
to phase segregate into tenorite and wurtzite phases below $T\sim$
900 to 1200 K, respectively.
Despite the simplicity of the nearest neighbor fixed lattice model,
the model describes the configurational landscape extremely well 
even without including the local lattice distortions 
arising from random cation mixing. 
Ultimately, this framework is flexible and should be adaptable to 
other structures and chemistries, thus enabling first-principles
based discovery of a wide range of entropy stabilized compounds.

\begin{suppinfo}
Detailed description of Monte Carlo method and supporting figures are included
as supplemental materials. 
List of all five component oxides (Table S1);
mixing enthalpy values of all two component oxides within four structures (Table S2);
antiferromagnetic type-II structure (Figure S1); 
four structural models considered in this study (Figure S2);
mixing enthalpy heat maps for wurtzite and zinc blende structures (Figure S3);
local enthalpy and entropy descriptors for four structures (Figure S4);
mixing enthalpy and entropy descriptors for four structures (Figure S5);
short range order parameters as a function of temperature for ESO (Figure S6);
local mixing enthalpy distribution for all five component combinations (Figure S7 and S8).
\end{suppinfo}

\begin{acknowledgement}
This work was supported by the LDRD Program of ORNL, managed by UT-Battelle,
LLC, for the U. S. DOE.
This research used resources of the Oak Ridge Leadership Facility, 
which is a DOE Office of Science User Facility supported under contact
DE-AC05-00OR22725 as well as National Energy Research Scientific 
Computing Center (NERSC), which is supported by the Office of Science 
of the U.S. Department of Energy under Contract No. DE-AC02-05CH11231.
KCP and VRC acknowledge James Morris for helpful discussions.
KCP acknowledges helpful discussions with German D. Samolyuk and Xianglin Liu.
\end{acknowledgement}

\bibliography{Bibliography}

\pagebreak

\begin{center}
\title{\LARGE\textbf{Supporting information: Predicting the phase stability of multi-component high entropy compounds}\par}
\end{center}
\appendix
\beginsupplement

\section{Metropolis Monte Carlo simulation details
\label{sec:SuppSec1}}
Utilizing the nearest neighbor model discussed in the main text, 
we employed Metropolis Monte Carlo simulations to study the mixing 
behaviour between different chemical species within the FCO at 
relevant temperatures ranging from $T=$ 300 to 1200 K.
Atomic configurations --- generated through randomly swapping unlike
atoms between different lattice sites in a $5\times5\times20$ periodic 
supercell (total 2000 cations) --- are sampled according to the Metropolis
criterion.
The simulation is started from a random atomic configuration.
%
%
A trial configuration is accepted according to the Boltzmann
probability, $p_\mathrm{B}$,
\begin{align}
\begin{split}
\label{eq:1}
p_\mathrm{B}=\mathrm{Min}\Bigg\{\mathrm{exp}\Bigg(\frac
{-\Delta H}
{k_\mathrm{B}T}\Bigg),1\Bigg\}
\end{split}
\end{align}
where, $\Delta H=(\Delta H^{n}_\mathrm{total}-\Delta H^{n-1}
_\mathrm{total})$ is the change in total mixing enthalpy between 
$n^\mathrm{th}$ and $n-1^\mathrm{th}$ steps, caused by swapping 
unlike chemical species.
$k_\mathrm{B}$ is Boltzmann constant and $T$ is absolute temperature.
The trial configuration is always accepted if $\Delta H \leq 0$.
However, if $\Delta H > 0$ the trial configuration is accepted 
by chance, according to the Boltzmann probability, $p_\mathrm{B}$.
In addition to the Metropolis sampling of atomic configurations,
we also assign the phase composition, by assigning a 
``global'' phase $P=\{\mathrm{G,R,T,W,Z}\},$ to the atomic configuration.
While $P=\mathrm{G}$ represents mixture of multiple phases in the 
lattice model, whose phase fractions are estimated using Eq. (7)
in the main text, $P=\{\mathrm{R,T,W,Z}\}$ represents pure 
rock salt, tenorite, wurtzite and Zinc blende phases, respectively.
Two schemes were tested to update the phase composition of the 
configuration, both resulting in approximately equal phase 
compositions and short range order parameter trends as a 
function of $T$. 
In the first scheme, the phase composition is updated after $N_p$
atomic updates, according to the metropolis criterion.
For example, the phase is updated from $P=\mathrm{G}$ to 
$P=\mathrm{R}$ according to the Boltzmann probability, $p_\mathrm{B}$,
where,
\begin{align}
\begin{split}
\label{eq:2}
\Delta H=\Delta H^n_\mathrm{total}[\mathrm{FCO},\mathrm{R}]-
\Delta H^n_\mathrm{total}[\mathrm{FCO},\mathrm{G}]),
\end{split}
\end{align}
is the energy difference between the configuration in pure rock salt
phase and the ground phase with mixed phase composition.
The results did not change significantly between $N_p=5$ and $2$.
In the second scheme, the phase is updated from the ground phase
to the pure phase at each step, according the Boltzmann probability,
where,
\begin{align}
\begin{split}
\label{eq:3}
\Delta H=\Delta H^{n}_\mathrm{total}[\mathrm{FCO},\mathrm{R}]-
\Delta H^{n-1}_\mathrm{total}[\mathrm{FCO},\mathrm{G}]),
\end{split}
\end{align}
measures the energy difference between the trail configuration in
the rock salt phase and the previous confuration in ground phase.
%
%
%
%
%
%
%
The Monte Carlo simulation is ran isothermally at several temperature 
values within a relevant temperature range.
The expectation values of phase fraction $\langle \phi(P) \rangle$
and short range order parameter $\langle \alpha_{AA'} \rangle$ are 
estimated for each simulation at constant temperature.

\begin{table}[H]
\renewcommand{\arraystretch}{1.2}
\caption{\label{tab:orderP}
List of five component combinations chosen from 
set of cations, $A$=\{Ca, Co, Cu, Fe, Mg, Mn, Ni, Zn\}.
}
\begin{tabular}{cc|cc}
\hline
Index & Combination & Index & Combination\\
\hline
1 	& CaCoCuFeMg 	& 29 	& CaCuMgNiZn\\
2 	& CaCoCuFeMn 	& 30 	& CaCuMnNiZn\\
3 	& CaCoCuFeNi 	& 31 	& CaFeMgMnNi\\
4 	& CaCoCuFeZn 	& 32 	& CaFeMgMnZn\\
5 	& CaCoCuMgMn 	& 33 	& CaFeMgNiZn\\
6 	& CaCoCuMgNi 	& 34 	& CaFeMnNiZn\\
7 	& CaCoCuMgZn 	& 35 	& CaMgMnNiZn\\
8 	& CaCoCuMnNi 	& 36 	& CoCuFeMgMn\\
9 	& CaCoCuMnZn 	& 37 	& CoCuFeMgNi\\
10 	& CaCoCuNiZn 	& 38 	& CoCuFeMgZn\\
11 	& CaCoFeMgMn 	& 39 	& CoCuFeMnNi\\
12 	& CaCoFeMgNi 	& 40 	& CoCuFeMnZn\\
13 	& CaCoFeMgZn 	& 41 	& CoCuFeNiZn\\
14 	& CaCoFeMnNi 	& 42 	& CoCuMgMnNi\\
15 	& CaCoFeMnZn 	& 43 	& CoCuMgMnZn\\
16 	& CaCoFeNiZn 	& 44 	& CoCuMgNiZn\\
17 	& CaCoMgMnNi 	& 45 	& CoCuMnNiZn\\
18 	& CaCoMgMnZn 	& 46 	& CoFeMgMnNi\\
19 	& CaCoMgNiZn 	& 47 	& CoFeMgMnZn\\
20 	& CaCoMnNiZn 	& 48 	& CoFeMgNiZn\\
21 	& CaCuFeMgMn 	& 49 	& CoFeMnNiZn\\
22 	& CaCuFeMgNi 	& 50 	& CoMgMnNiZn\\
23 	& CaCuFeMgZn 	& 51 	& CuFeMgMnNi\\
24 	& CaCuFeMnNi 	& 52 	& CuFeMgMnZn\\
25 	& CaCuFeMnZn 	& 53 	& CuFeMgNiZn\\
26 	& CaCuFeNiZn 	& 54 	& CuFeMnNiZn\\
27 	& CaCuMgMnNi 	& 55 	& CuMgMnNiZn\\
28 	& CaCuMgMnZn 	& 56 	& FeMgMnNiZn\\
\hline
\end{tabular}
\end{table}

\begin{longtable}{cccccccc}
\caption{Mixing enthalpy values of two component oxides computed from DFT in eV/$A-$site.} \label{tab:long} \\

\hline \multicolumn{1}{c}{Index} & \multicolumn{1}{c}{TCO} & \multicolumn{1}{c}{Rock salt} & \multicolumn{1}{c}{Tenorite} & \multicolumn{1}{c}{Wurtzite} & \multicolumn{1}{c}{Zinc blende} & \multicolumn{1}{c}{Ground phase} \\ \hline 
\endfirsthead

\multicolumn{8}{c}%
{{\bfseries \tablename\ \thetable{} -- continued from previous page}} \\
\hline \multicolumn{1}{c}{Index} & \multicolumn{1}{c}{TCO} & \multicolumn{1}{c}{Rock salt} & \multicolumn{1}{c}{Tenorite} & \multicolumn{1}{c}{Wurtzite} & \multicolumn{1}{c}{Zinc blende} & \multicolumn{1}{c}{Ground phase} \\ \hline 
\endhead

\hline \multicolumn{8}{r}{{Continued on next page}} \\ \hline
\endfoot

\hline
\endlastfoot
1 & Ca-Ca & 0.000 & 0.000 & 0.015 & 0.038 & Rock salt \\
2 & Ca-Co & 0.027 & 0.025 & 0.033 & 0.039 & Tenorite \\
3 & Ca-Cu & 0.014 & 0.000 & 0.029 & 0.041 & Tenorite \\
4 & Ca-Fe & 0.055 & 0.053 & 0.066 & 0.044 & Zinc blende \\
5 & Ca-Mg & 0.024 & 0.024 & 0.039 & 0.045 & Rock salt \\
6 & Ca-Mn & 0.021 & 0.021 & 0.033 & 0.045 & Rock salt \\
7 & Ca-Ni & 0.031 & 0.035 & 0.035 & 0.034 & Rock salt \\
8 & Ca-Zn & 0.018 & 0.021 & 0.027 & 0.031 & Rock salt \\
9 & Co-Co & 0.000 & 0.039 & 0.009 & 0.011 & Rock salt \\
10 & Co-Cu & 0.005 & 0.007 & 0.054 & 0.052 & Rock salt \\
11 & Co-Fe & 0.073 & 0.042 & 0.055 & 0.058 & Tenorite \\
12 & Co-Mg & 0.003 & 0.048 & 0.014 & 0.019 & Rock salt \\
13 & Co-Mn & 0.019 & 0.020 & 0.029 & 0.030 & Rock salt \\
14 & Co-Ni & 0.002 & 0.044 & 0.031 & 0.033 & Rock salt \\
15 & Co-Zn & 0.014 & 0.042 & 0.005 & 0.006 & Wurtzite \\
16 & Cu-Cu & 0.013 & 0.000 & 0.057 & 0.052 & Tenorite \\
17 & Cu-Fe & 0.053 & 0.043 & 0.072 & 0.068 & Tenorite \\
18 & Cu-Mg & 0.004 & 0.002 & 0.039 & 0.040 & Tenorite \\
19 & Cu-Mn & 0.022 & 0.020 & 0.056 & 0.057 & Tenorite \\
20 & Cu-Ni & 0.007 & 0.006 & 0.055 & 0.056 & Tenorite \\
21 & Cu-Zn & 0.015 & 0.012 & 0.028 & 0.029 & Tenorite \\
22 & Fe-Fe & 0.082 & 0.103 & 0.092 & 0.082 & Rock salt \\
23 & Fe-Mg & 0.043 & 0.041 & 0.072 & 0.074 & Tenorite \\
24 & Fe-Mn & 0.056 & 0.058 & 0.066 & 0.084 & Rock salt \\
25 & Fe-Ni & 0.046 & 0.046 & 0.067 & 0.044 & Zinc blende \\
26 & Fe-Zn & 0.051 & 0.051 & 0.062 & 0.063 & Rock salt \\
27 & Mg-Mg & -0.000 & 0.068 & 0.020 & 0.026 & Rock salt \\
28 & Mg-Mn & 0.020 & 0.020 & 0.033 & 0.036 & Rock salt \\
29 & Mg-Ni & -0.002 & 0.068 & 0.003 & 0.069 & Rock salt \\
30 & Mg-Zn & 0.007 & 0.057 & 0.009 & 0.012 & Rock salt \\
31 & Mn-Mn & 0.030 & 0.029 & 0.043 & 0.045 & Tenorite \\
32 & Mn-Ni & 0.026 & 0.026 & 0.030 & 0.028 & Rock salt \\
33 & Mn-Zn & 0.026 & 0.026 & 0.024 & 0.025 & Wurtzite \\
34 & Ni-Ni & 0.000 & 0.052 & 0.010 & 0.056 & Rock salt \\
35 & Ni-Zn & 0.012 & 0.050 & 0.026 & 0.028 & Rock salt \\
36 & Zn-Zn & 0.018 & 0.048 & 0.000 & 0.001 & Wurtzite \\
\end{longtable}

\begin{figure*}[b]
\includegraphics[width=0.5\columnwidth]{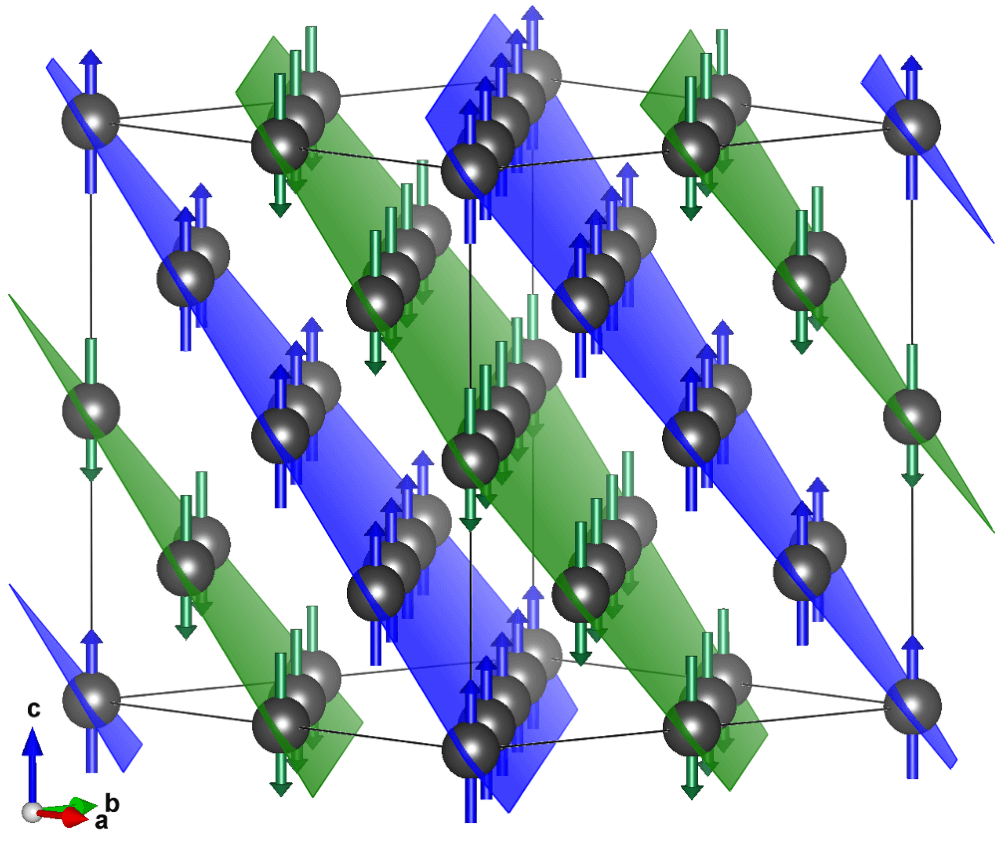}
\caption{\label{fig:AFM}
AFM-II magnetic structure represented in a 2x2x2 rock salt super cell.
Alternate blue and green (111) lattice planes present the spin-up and
spin-down planes, respectively.
}
\end{figure*}

\begin{figure*}[tbh!]
\includegraphics[width=1\columnwidth]{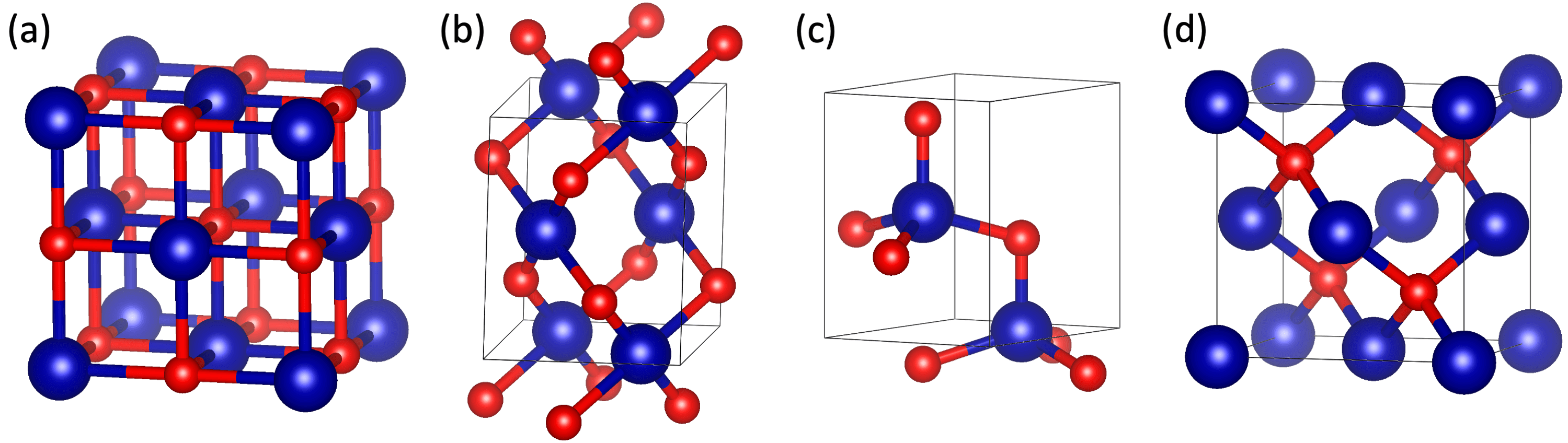}
\caption{\label{fig:structure}
Structural models for \textbf{(a)} rock salt, \textbf{(b)} tenorite,
\textbf{(c)} wurtzite, and \textbf{(d)} zinc belnde.
Red spheres represent oxygen and blue atoms represent
one of the $A$-site cations considered in this study,
$A=$\{Ca,Co,Cu,Fe,Mg,Mn,Ni,Zn\}
}
\end{figure*}

\begin{figure*}[tbh!]
\includegraphics[width=1\columnwidth]{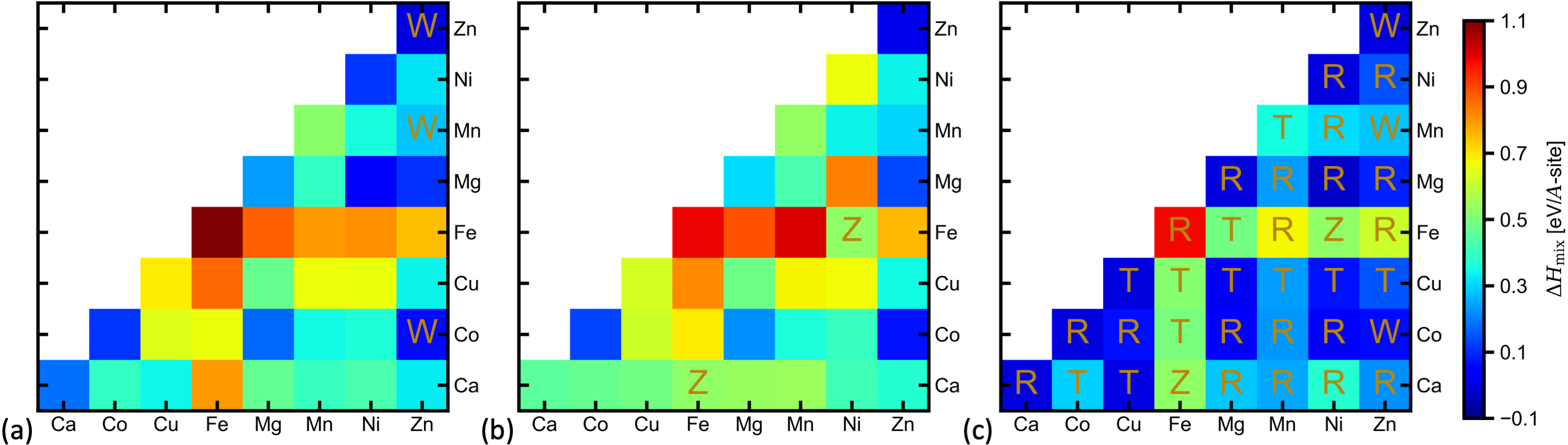}
\caption{\label{fig:mixingE}
Heat maps for mixing enthalpy in eV/$A$-site of $(AA^{'})$O ternary 
oxides in phase $P$, $E_\mathrm{M}((AA^{'})\mathrm{O},P)$.
The mixing enthalpies are shown for \textbf{(a)} wurtzite, 
\textbf{(b)} zinc belnde, and \textbf{(c)} ground phases.
\textbf{(c)} Also shows the annotated ground phase for each $A-A^{'}$
combination.
Only the lower triangular portion of the symmetric data is shown in 
all panels to avoid confusion.
The color legend represents the mixing enthalpy in [eV$/A$-site] --- 
blue and red represents low ($\sim$ -0.1) and high ($\sim$ 1.1) bond 
energies, respectively.
}
\end{figure*}

\begin{figure*}[t]
\includegraphics[width=1\columnwidth]{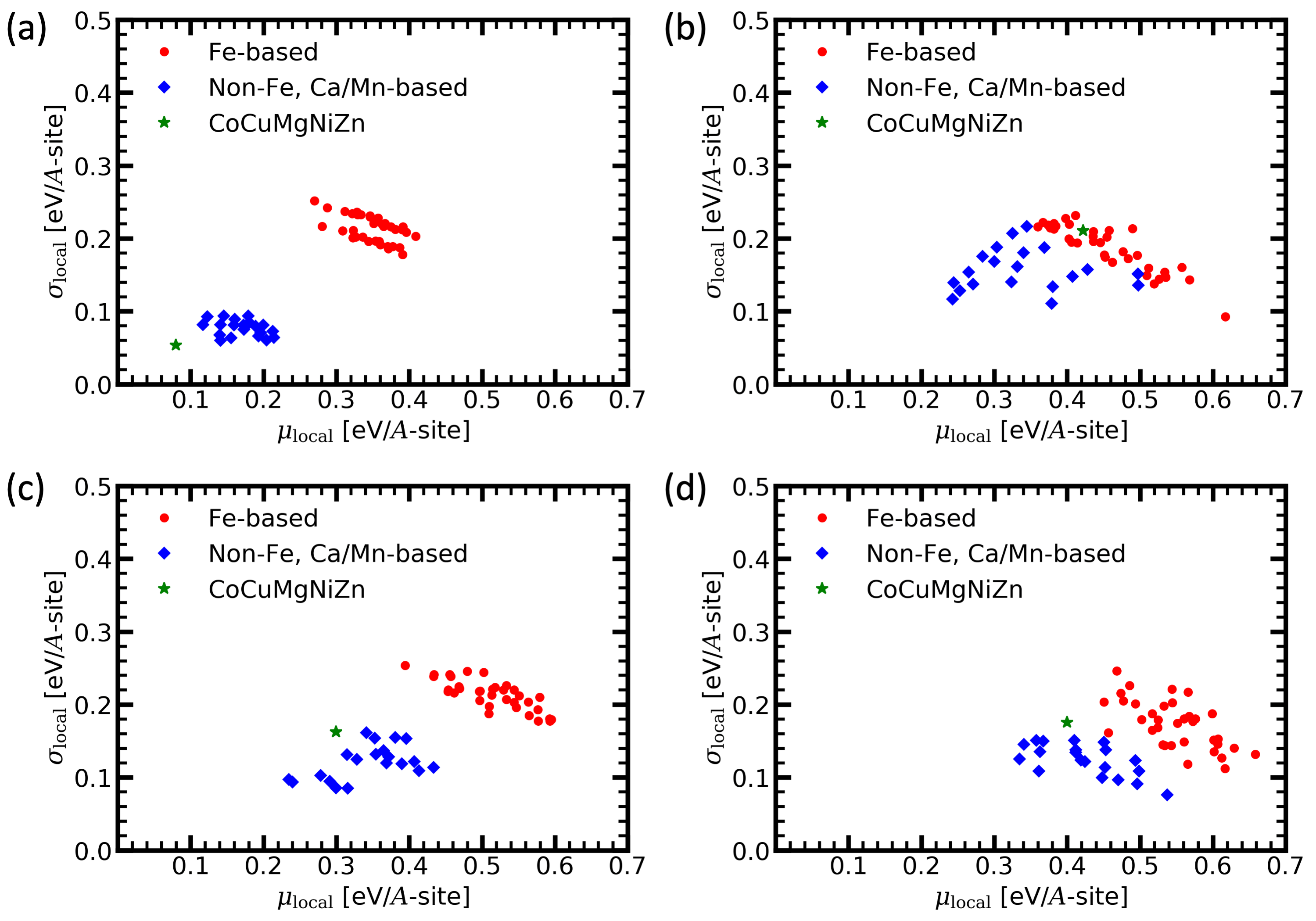}
\caption{\label{fig:localMu}
Comparison of enthalpy and entropy indicators, $\mu_\mathrm{local}$ and 
$\sigma_\mathrm{local}$, respectively, for all 56 FCOs in 
\textbf{(a)} rock salt, \textbf{(b)} tenorite, \textbf{(c)} wurtzite, 
and \textbf{(d)} zinc blende structures.
(1) Fe-based FCO cluster is shown in red, 
(2) Ca- or Mn-based (without Fe) FCO cluster is shown in blue, and 
(3) The entropy stabilized oxide --- (CoCuMgNiZn)O --- shown in green,
is unique FCO with least-$\mu_\mathrm{local}$ and
least-$\sigma_\mathrm{local}$ in rock salt structure. 
}
\end{figure*}

\begin{figure*}[t]
\includegraphics[width=1\columnwidth]{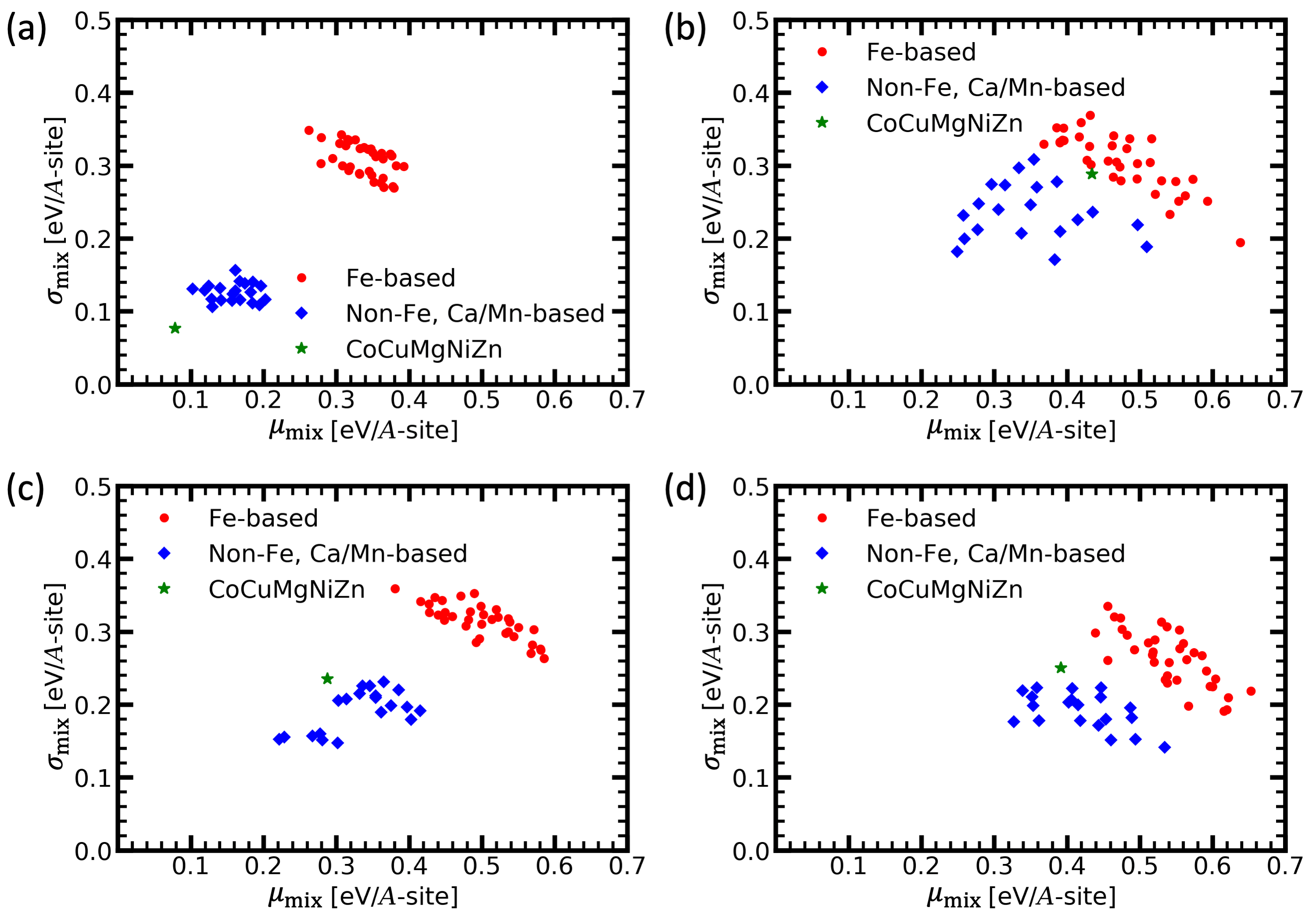}
\caption{\label{fig:mixingMu}
Comparison of enthalpy and entropy indicators, $\mu_\mathrm{M}$ and 
$\sigma_\mathrm{M}$, respectively, for all 56 FCOs in 
\textbf{(a)} rock salt, \textbf{(b)} tenorite, 
\textbf{(c)} wurtzite, and \textbf{(d)} zinc blende structures.
(1) Fe-based FCO cluster is shown in red circles, 
(2) Ca- or Mn-based (without Fe) FCO cluster is shown in blue diamonds,
and (3) The entropy stabilized oxide --- (CoCuMgNiZn)O --- shown in
green star, is unique FCO with least-$\mu_\mathrm{local}$ and least-$\sigma_\mathrm{local}$ in rock salt structure. 
}
\end{figure*}

\begin{figure*}[t]
\includegraphics[width=1\columnwidth]{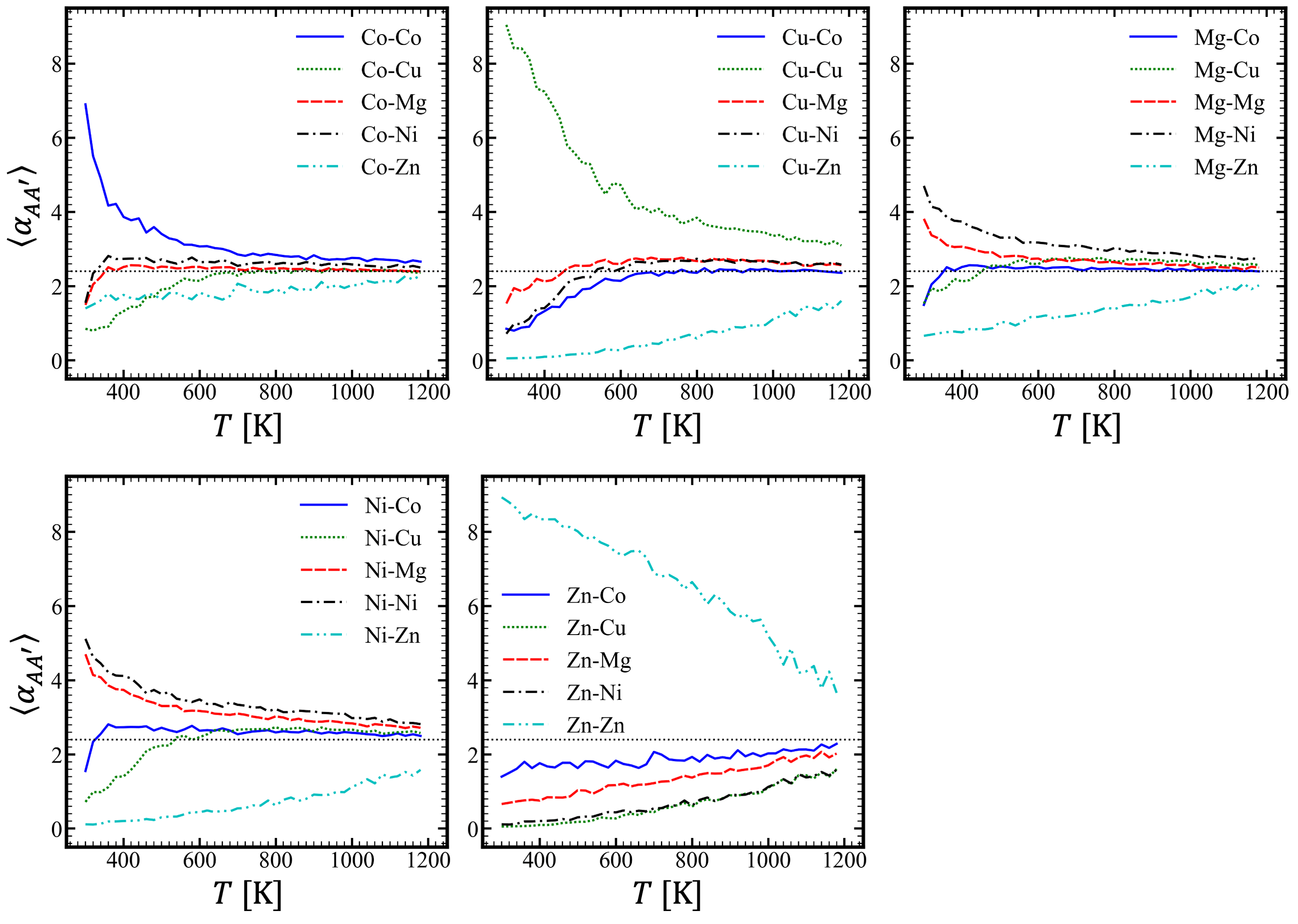}
\caption{\label{fig:sopall}
Short range order parameter plotted for all $AA^{'}$ combinations, as
a function of temperature for the (CoCuMgNiZn)O.
}
\end{figure*}

\begin{figure*}[hp]
\includegraphics[width=0.97\columnwidth]{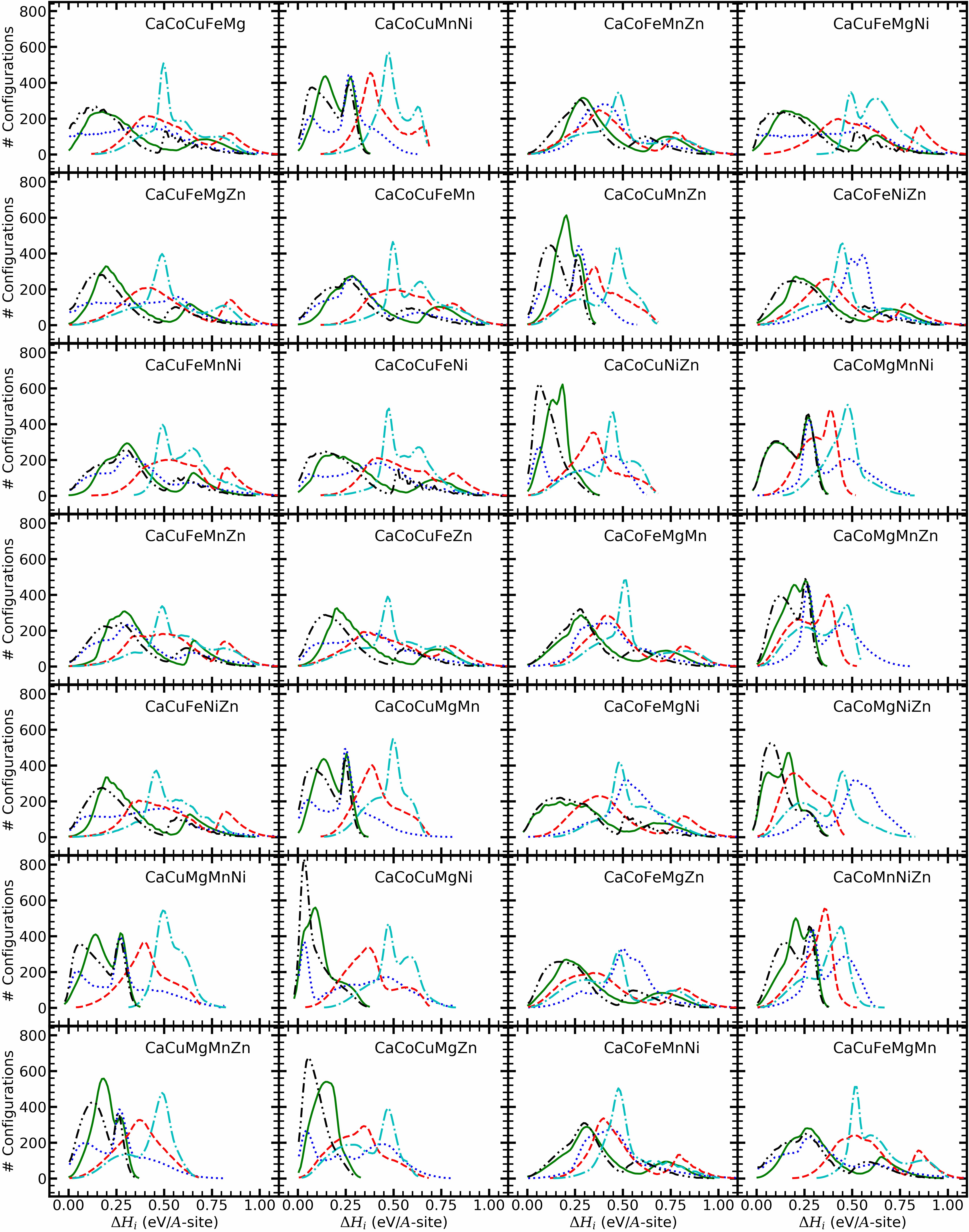}
\caption{\label{fig:cdos1}
The distribution of the local mixing enthalpies of all possible local
configurations for five component oxides (1 to 28), plotted
in separate panel.
Solid green, dotted blue, dashed red, dot-dashed cyan and black
dot-dot-dashed lines represent distributions for rock salt, tenorite,
wurtizite, zinc blende and ground phases, respectively.
}
\end{figure*}

\begin{figure*}[hp]
\includegraphics[width=0.97\columnwidth]{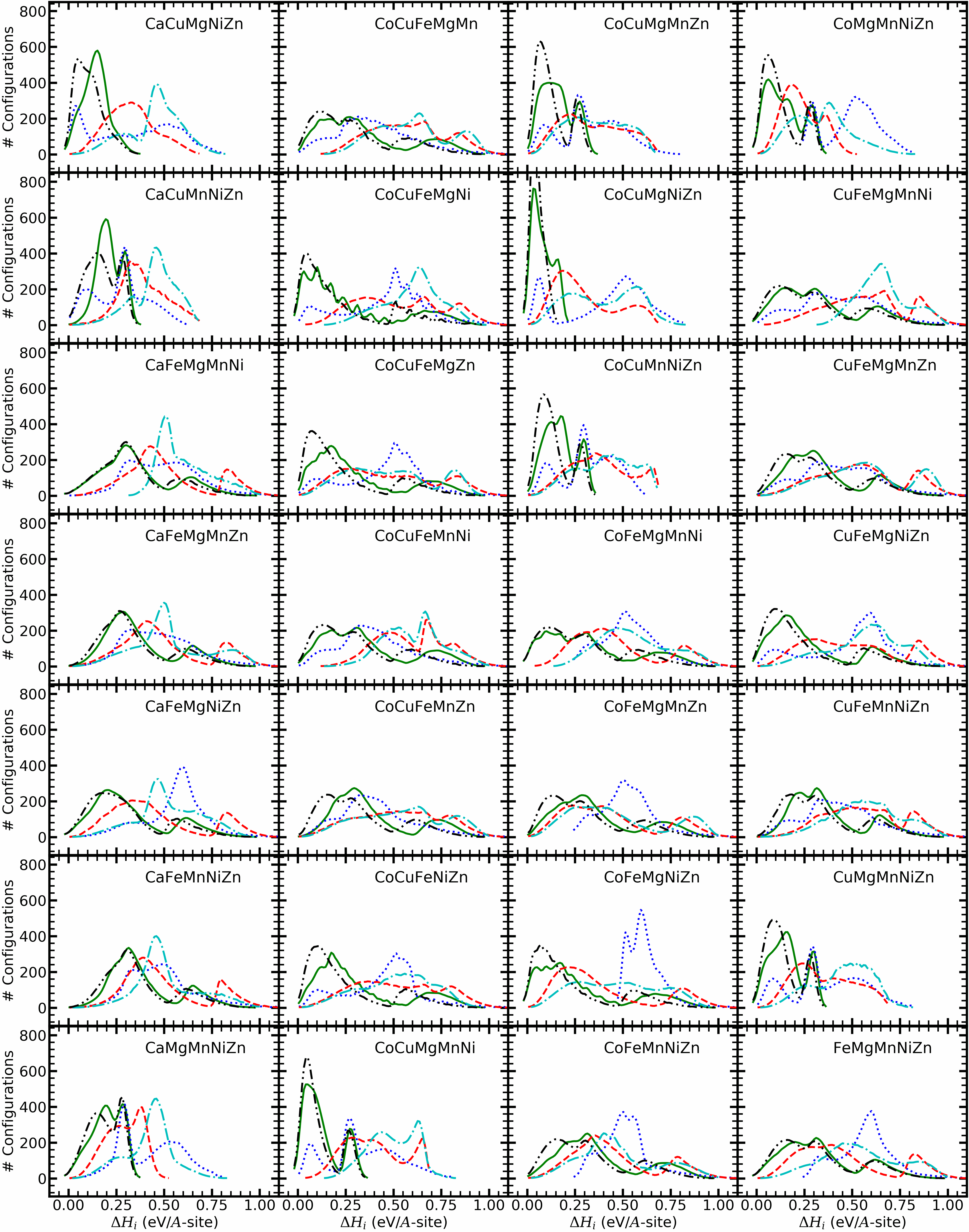}
\caption{\label{fig:cdos2}
The distribution of the local mixing enthalpies of all possible local
configurations for five component oxides (29 to 56), plotted
in separate panel.
Solid green, dotted blue, dashed red, dot-dashed cyan and black
dot-dot-dashed lines represent distributions for rock salt, tenorite,
wurtizite, zinc blende and ground phases, respectively.
}
\end{figure*}

\end{document}